\documentclass[twocolumn, superscriptaddress,amsmath,amssymb,aps,pra,
showkeys,showpacs,nofootinbib] {revtex4-1}
\usepackage{graphicx}
\usepackage{color}
\usepackage{bm}
\usepackage{hyperref}
\usepackage{dcolumn}
\usepackage{slashed}

\begin{document}
\title{Continuous Generation and Stabilization of Mesoscopic Field Superposition States in a Quantum Circuit}
\author{Ananda Roy}
\email{ananda.roy@yale.edu}
\affiliation{Department of Applied Physics, Yale University, PO BOX 208284, New Haven, CT 06511}
\affiliation{Laboratoire Pierre Aigrain, Ecole Normale Supérieure, CNRS (UMR 8551),
Université P. et M. Curie, Université D. Diderot 
24, rue Lhomond, 75231 Paris Cedex 05, France}
\affiliation{INRIA Paris-Rocquencourt, Domaine de Voluceau, B.P. 105, 78153 Le Chesnay Cedex, France}
\author{Zaki Leghtas}
\affiliation{Department of Applied Physics, Yale University, PO BOX 208284, New Haven, CT 06511}
\author{A. Douglas Stone}
\affiliation{Department of Applied Physics, Yale University, PO BOX 208284, New Haven, CT 06511}
\author{Michel Devoret}
\affiliation{Department of Applied Physics, Yale University, PO BOX 208284, New Haven, CT 06511}
\author{Mazyar Mirrahimi}
\affiliation{INRIA Paris-Rocquencourt, Domaine de Voluceau, B.P. 105, 78153 Le Chesnay Cedex, France}
\affiliation{Department of Applied Physics, Yale University, PO BOX 208284, New Haven, CT 06511}

\begin{abstract}
While dissipation is widely considered as being harmful for quantum coherence, it can, when properly engineered, lead to the stabilization of non-trivial pure quantum states. We propose a scheme for continuous generation and stabilization of  Schr\"{o}dinger cat states in a cavity using dissipation engineering. We first generate non-classical photon states with definite parity by 
means of a two-photon drive and dissipation, and then stabilize these transient states against single-photon decay.  
The single-photon stabilization is autonomous, and is implemented through a second engineered bath, which exploits
the photon number dependent frequency-splitting due to Kerr interactions in the strongly dispersive regime of circuit QED.  Starting with the Hamiltonian of the baths plus cavity, we derive an effective model of only the cavity photon states along with analytic expressions for relevant physical quantities, such as the stabilization rate. The deterministic generation of such cat states is one of the key ingredients in performing universal quantum computation. 
\end{abstract}
\maketitle 

\section{Introduction}
\label{intro}
Quantum computing has shown great promise as a resource providing exponential speedup over certain classical algorithms and as an indispensable tool for efficient simulation of quantum systems \cite{Deutsch_1985, Feynman_1985, Shor_1994, DiVincenzo_1995}. Recent years have seen considerable effort in understanding how a quantum computer outperforms its classical counterpart. An essential ingredient has been identified for systems performing universal quantum computation with continuous variables (e.g. modes of electromagnetic field): non-classical states, i.e. states displaying negativity in their Wigner function \cite{Bartlett_Nemoto_2002, Eisert_Plenio_2002, Fiurasek_2002, Giedke_Cirac_2002, Mari_Eisert_2012}. This can be achieved by engineering a Hamiltonian with terms higher than quadratic in mode-amplitude, for instance the Kerr Hamiltonian, which is quartic~\cite{Yurke_Stoler_1986}. Such a Hamiltonian, together with linear scattering elements like beam-splitters, drives and squeezers, is sufficient to perform arbitrary polynomial transformations of the mode variables \cite{Lloyd_Braunstein_1999}. 

Non-classical input states such as single photons and superpositions of coherent states are the main candidates for universal quantum computation with linear optical circuits \cite{KLM_2001, Jeong_Kim_2002, Ralph_Milburn_2002, Kok_Milburn_2007}. This has stimulated experiments in which single-photon states are generated in a heralded \cite{Lvovsky_Schiller_2001, Fasel_Zbinden_2004, Neergaard-Nielsen_2007} and on demand \cite{Yoshikawa_Furusawa_2013, Bimbard_Grangier_2014} manner.  Various experimental schemes have likewise  produced and observed superposition of coherent states in optical systems in a heralded manner using photon subtraction \cite{Ourjoumtsev_Grangier_2006, Neergaard-Nielsen_2006, Wakui_Sasaki_2007, Ourjoumtsev_Grangier_2007, Hiroki_Sasaki_2008}.  In the context of cavity/circuit QED, such superposition states have been generated by mapping a qubit state to a coherent state superposition in a heralded manner~\cite{Delaglise_Haroche_2008} and on demand~\cite{Vlastakis_Schoelkopf_2013}. Here, we go a step further and we address the question of robustly stabilizing cavity photons in a superposition of coherent states. This could act as a continuous and deterministic source of non-classical input states in quantum information processing protocols.

To that end, we apply a dissipation engineering technique leading to an autonomous preparation and protection against decoherence of these states~\cite{Poyatos_Zoller_1996}. An earlier theoretical proposal within the framework of  cavity QED with Rydberg atoms describe such a stabilization by an adequate engineered system-bath interaction \cite{Sarlette_Rouchon_2011}. The current proposal is adapted to photon states in quantum superconducting circuits, and requires only the application of continuous-wave (CW) microwave drives of fixed frequencies and amplitudes, thus greatly simplifying an experimental implementation. 

The first stage of our proposal builds on recent theoretical work in such systems~\cite{Mirrahimi_Devoret_2014}
in which a bath was engineered such that photons are only exchanged in pairs.
Such a nonlinear system-bath interaction was shown to stabilize the manifold spanned by two coherent states $|\alpha\rangle$ and $|-\alpha\rangle$ (where $\alpha$, the coherent state amplitude parameter, is determined by a tunable external drive). Very recently this proposal has been implemented successfully in an experimental set-up \cite{Zaki}.  The dynamics generated by such an interaction conserves photon number parity: an initial vacuum state $|0\rangle$ would therefore converge to the even Schr\"{o}dinger cat state  $|C_\alpha^+\rangle  = \sum_{n=0}^{\infty}c_{2n}|2n\rangle$,  $c_{m} = \frac{e^{-|\alpha|^2/2}}{\sqrt{2(1+e^{-2|\alpha|^2})}}\frac{\alpha^{m}}{\sqrt{m!}}$. Similarly, an odd parity initial state will converge to the odd Schr\"{o}dinger cat state $|C_\alpha^-\rangle  = \sum_{n=0}^{\infty}c_{2n+1}|2n+1\rangle$. Finally, an initial state with undefined parity will converge to a final state of undefined parity. In practice however, while one can add a two-photon bath interaction which transiently dominates the dynamics, there will always be a residual single-photon loss channel that will decohere these parity superpositions, leading to a statistical mixture of $|\alpha\rangle$ and $|-\alpha\rangle$ in the steady-state.~\footnote{The changes in photon-number-parity resulting from single photon loss can, in principle, be continuously monitored \cite{Sun_Schoelkopf_2014} and compensated for, in a measurement based feedback scheme.}

In this paper we present a theoretical proposal where we autonomously compensate for single photon loss and ensure the stabilization of a single superposition  (e.g. $|C_{\alpha}^+\rangle = \frac{1}{\sqrt{2(1+e^{-2|\alpha|^2})}}(|\alpha\rangle + |-\alpha\rangle)$) in this manifold. Similarly to some recent autonomous stabilization protocols for superconducting qubits \cite{Geerlings_Devoret_2013, Shankar_Devoret_2013, Leghtas_Mirrahimi_2013}, we benefit from the high quality factors of the superconducting microwave resonators in presence of strong nonlinear interactions provided by Josephson elements. More precisely, we make use of dispersive (cross-Kerr) interaction  between two cavity modes mediated by a transmon qubit coupled to both of them~\cite{Eric}. Working in the strong dispersive regime~\cite{Schuster_Schoelkopf_2007}, we design an effective decay of the cavity mode from a cat state of odd parity to a cat state of even parity. This dissipation, together with the two-photon process, reduces the steady state from a manifold spanned by $\big\{|C_\alpha^+\rangle, |C_\alpha^-\rangle\big\}$ to a unique state ($|C_\alpha^+\rangle$). The full system requires only a high Q ``storage cavity", coupled to two low-Q ``readout cavities'' through Josephson junctions and requires cavity decay and coupling parameters well within the reach of current technology.  A trivial modification of the scheme leads to stabilization of $|C_\alpha^-\rangle$. Note that even though we use the term ``readout'' to refer to the dissipative baths, the information leaking through the ports associated with the two low-Q cavities does not need to be monitored. It suffices that it never returns to the stabilized ``storage cavity''. 

The paper is organized as follows: in Section \ref{Outline_Stabilization_Scheme}, we describe our dissipation engineering scheme that stabilizes an even Schr\"{o}dinger cat state. In Section \ref{Experimental_Implementation}, we describe the possible experimental implementation, engineering the Hamiltonian interactions and dissipation, that realizes the stabilization scheme. We sweep the parameters that are in principle tunable in an ongoing experiment to determine the optimal choice. Next, we perform adiabatic elimination of the faster dynamical variables to arrive at an effective interaction and dissipation for the storage cavity alone, providing analytic expressions for the various decay and interaction rates  (Sec. \ref{adel}). We summarize our results in Sec. \ref{concl}.

\section{Two-photon process and Parity Selection}
\label{Outline_Stabilization_Scheme}
In this section, we briefly outline the interaction and dissipation scheme that gives rise to an even Schr\"{o}dinger cat state ($|C_\alpha^+\rangle$) in the steady state regime. We assume, for the storage cavity, the existence of a single-photon decay channel which is the natural dominant decoherence channel in the absence of engineered system-bath interactions.  We further assume that we have engineered two additional decay channels: the two-photon decay channel through which pairs of photons are lost into the environment (following previous work \cite{Wolinsky_Carmichael_1988, Hach_Gerry_1994, Gilles_Knight_1994, Krippner_Reid_1994, Mirrahimi_Devoret_2014}), and a new, parity-selection decay channel, which leads to an effective transfer of population from the odd to the even photon number parity manifold. These decay channels are characterized by effective decay rates $\kappa_{\rm{2ph}}$ and $\kappa_{\rm{ps}}$, respectively, and we assume that we can engineer them to be much larger  
than the rate of single photon loss ($\kappa_{\rm{1ph}}$) for the relevant cavity modes:
\begin{equation}
 \label{timescale}
 \kappa_{\rm{1ph}}\ll\kappa_{\rm{2ph}}, \kappa_{\rm{ps}}. 
\end{equation}

\subsection{Two-Photon Process}
Consider a cavity mode coupled to a bath and a drive such that it absorbs or loses photons only in pairs. Denoting the annihilation operator for this two-photon driven-dissipative harmonic oscillator as $\mathbf{a_s}$, the master equation for the mode is: 
\begin{eqnarray}
\label{w/oerror}
 \frac{d\rho}{dt} &=& -i\big[\mathbf{H}_{\rm{2ph}},\rho\big] + \kappa_{\rm{2ph}}{\cal D}(\mathbf{a_s}^2)\rho + \kappa_{\rm{1ph}}{\cal D}(\mathbf{a_s})\rho,
\end{eqnarray}
where ${\cal D}(\hat{O})\rho = \hat{O}\rho\hat{O}^\dagger - \frac{1}{2}\hat{O}^\dagger\hat{O}\rho - \frac{1}{2}\rho\hat{O}^\dagger\hat{O}$ is the usual Lindblad operator, $\mathbf{H}_{\rm{2ph}}=i\big(\epsilon_{\rm{2ph}}\mathbf{a_s}^{\dagger2} - \epsilon_{\rm{2ph}}^*\mathbf{a_s}^2\big)$ and $\epsilon_{\rm{2ph}}$ is the two-photon drive strength.
As noted, for  $\kappa_{\rm{1ph}} = 0$, one can show that starting from vacuum ($\rho(t=0)=|0\rangle\langle0|$), the density matrix converges towards $\rho(t\rightarrow\infty) = |C_\alpha^+\rangle\langle C_\alpha^+|$, where $\alpha = \sqrt{2\epsilon_{\rm{2ph}}/\kappa_{\rm{2ph}}}$ \cite{Mirrahimi_Devoret_2014}. In the presence of single photon loss, due to the random photon jumps, the cat state undergoes decoherence resulting in an incoherent mixture of $|\alpha\rangle$ and $|-\alpha\rangle$. 
 
\subsection{Parity Selection} 
\begin{figure}[!h]
\centering
 \includegraphics[width = 0.5\textwidth]{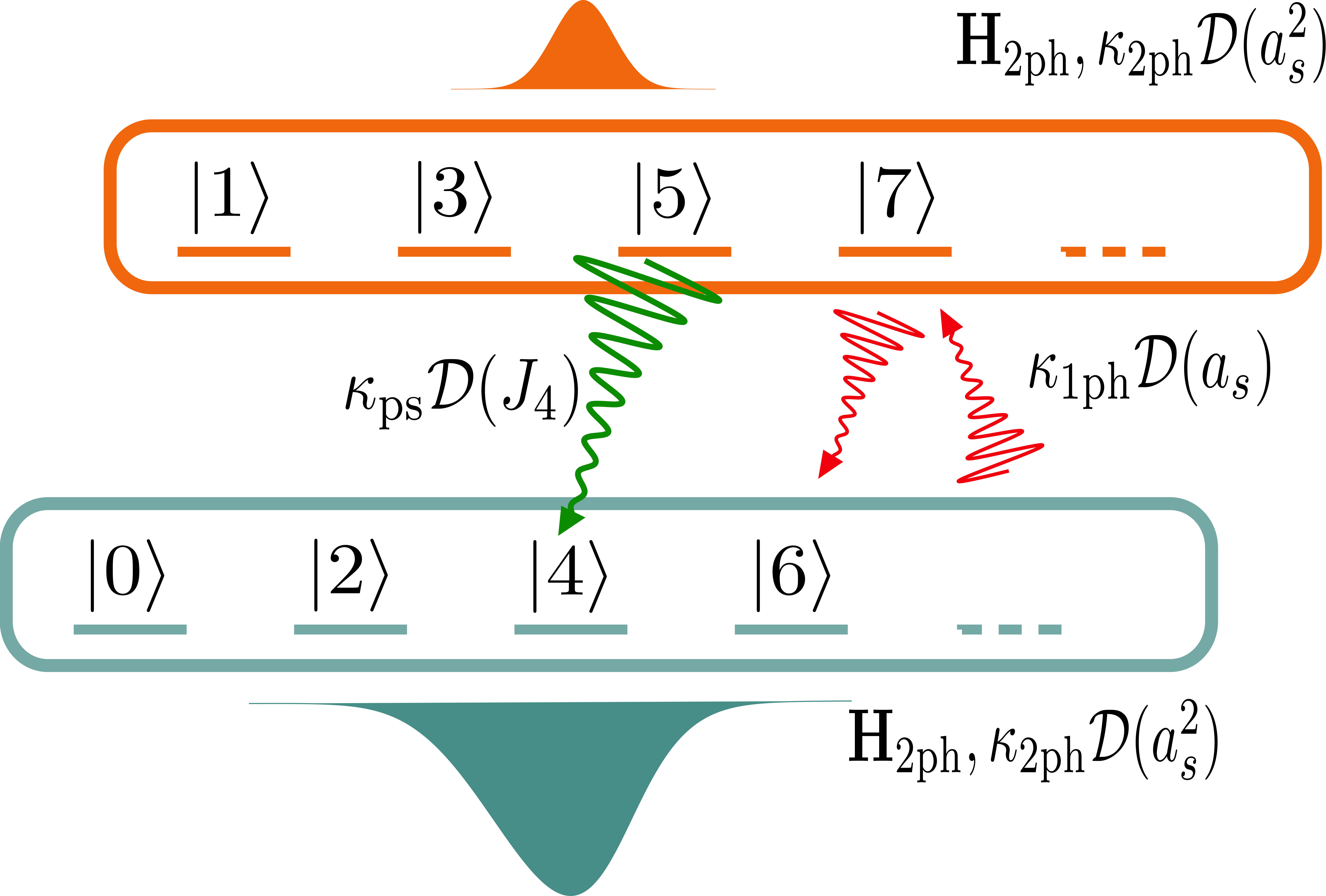}
\caption{\label{manifold} Schematic for the stabilization of the ``cat state", $|C_{\alpha = 2}^+\rangle$. The two-photon drive and dissipation (denoted by ${\mathbf{H}}_{\rm{2ph}}, \kappa_{\rm{2ph}}{\cal D}(a_s^2)$) act on the even and odd manifolds, shown in blue and orange respectively. In the absence of single photon loss and starting from vacuum, the odd manifold remains unpopulated, while the even manifold population is distributed to realize an even cat state. However, single photon loss (shown in red) denoted by $\kappa_{\rm{1ph}}{\cal D}(a_s)$ transfers some of the population to the odd manifold, where it is distributed as in an odd cat state due to the two-photon drive/dissipation. We propose to engineer a dissipation interaction from $|5\rangle$ to $|4\rangle$ (in green) denoted by $\kappa_{\rm{ps}}{\cal D}(J_4)$. This dissipation, together with the two-photon process, transfers excitations from the odd to even manifold and stabilizes the desired cat state.}
\end{figure}

In order to compensate for the decoherence due to single photon loss, we consider the action of effective jump operators of the form $\mathbf{J}_{2n} = |2n\rangle\langle 2n+1|$, which acting on the odd number states bring it to the immediate lower even number state. \footnote{If the desired target state is $|C_\alpha^-\rangle$, one needs to consider jump operators of the form $\mathbf{J}_{2n-1} = |2n-1\rangle\langle 2n|$.} This transfers the excitations from the odd parity manifold, which gets populated due to single photon loss, to the even parity manifold. Once the population is transferred to the even manifold, the two-photon process redistributes the population over the even manifold so as to reach the steady-state determined by the two-photon bath plus drive, $|C_\alpha^+\rangle$. Let us consider, for simplicity, only one such operator: $\mathbf{J}_{2\tilde{n}} = |2\tilde{n}\rangle\langle2\tilde{n}+1|$, where $2\tilde{n}$ is the integer closest to the average number of photons in the even cat $|C_\alpha^+\rangle$.  The two-photon process acts also on the odd manifold, where it redistributes population, with maximum around $|2\tilde{n}+1\rangle$, so as to funnel probability density towards the escape channel given by the jump operator, $\mathbf{J}_{2\tilde{n}}$. Thus, although by itself this jump operator only transfers the population from the Fock state $|2\tilde{n}+1\rangle$ to $|2\tilde{n}\rangle$, together with the two-photon process, it drains the population from the odd to the even manifold (cf. Fig.~\ref{manifold}). The rate associated with this parity selection process will be denoted by $\kappa_{\rm{ps}}$.  Thus, we can write down the master equation governing the stabilized evolution of the cavity mode: 
\begin{eqnarray}
\label{werrorcorr}
 \frac{d\rho}{dt} &=& -i\big[\mathbf{H}_{\rm{2ph}},\rho\big] + \kappa_{\rm{2ph}}{\cal D}(\mathbf{a_s}^2)\rho + \kappa_{\rm{1ph}}{\cal D}(\mathbf{a_s})\rho\nonumber\\&& + \kappa_{\rm{ps}}{\cal D}(\mathbf{J}_{2\tilde{n}})\rho. 
\end{eqnarray}

\begin{figure}[!h]
\centering
 \includegraphics[height = 4.5cm , width = 9.2cm]{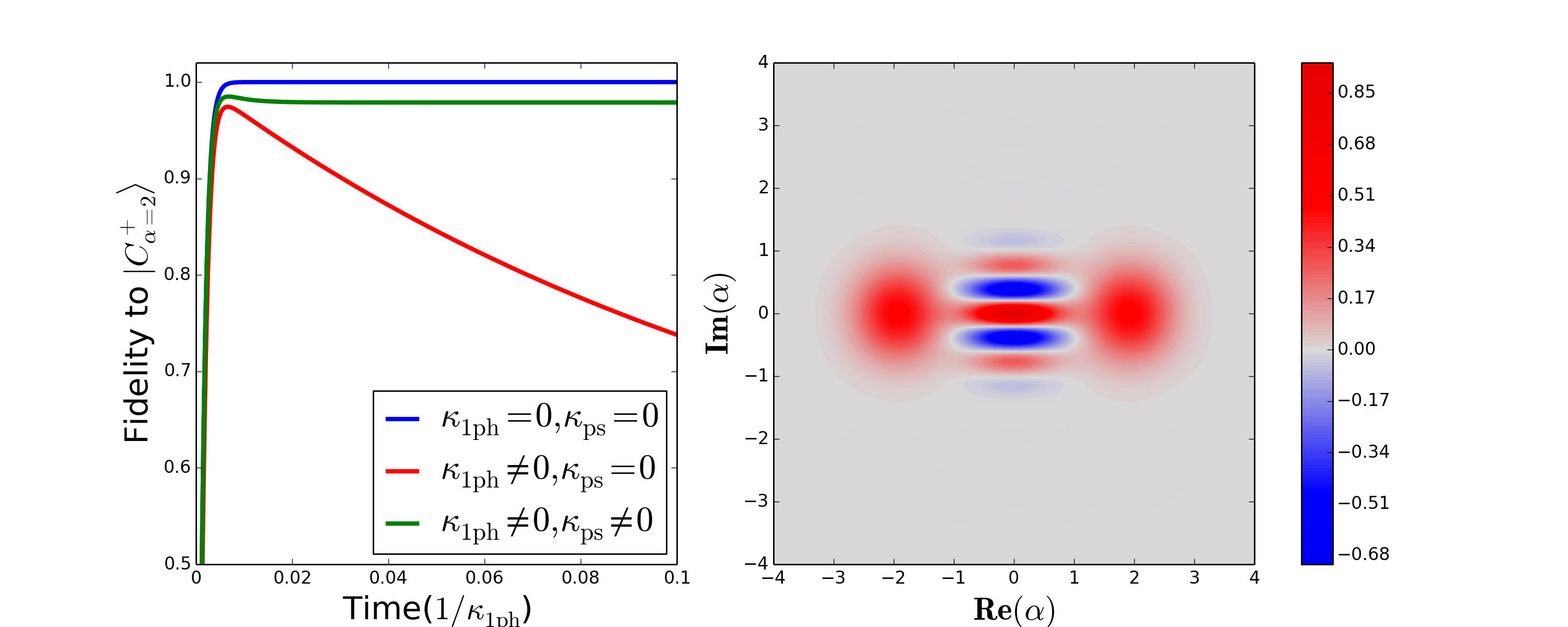}
\caption{\label{corrcomp2} Fidelity with respect to the target state $|C_{\alpha = 2}^+\rangle$  (left panel) and Wigner function of the steady-state (right panel). The parity-selecting dissipation and the two-photon dissipation/drive, in the presence of single photon loss, stabilizes the even cat state. The dissipation rates are $\kappa_{\rm{2ph}} = 250 \kappa_{\rm{1ph}}, \kappa_{\rm{ps}} = 760 \kappa_{\rm{1ph}}$. The evolution of fidelity is shown in absence of single photon loss (blue), presence of single photon loss and absence of parity selection (red), and lastly, in presence of single photon loss and parity selection (green). Steady state Wigner function of stabilized cat state is shown in presence of single and two-photon loss and parity selection.}
\end{figure}

In Fig.~\ref{corrcomp2}, we show the results of simulation of this equation. On the right is shown the Wigner function for the final state for $\alpha=2$ when all terms are present in the evolution equation.  The interference fringes near the origin clearly show the negativity of the Wigner function. On the left we show the time evolution of the fidelity of the solution of the evolution equation with respect to the ideal target state for three cases.  In the absence of single-photon loss 
($\kappa_{\rm{1ph}}=0$), the fidelity approaches unity at a rate determined by $\kappa_{\rm{2ph}}$.  When single-photon loss is added but not stabilization ($\kappa_{\rm{1ph}} \neq 0, \kappa_{\rm{ps}} = 0$) the fidelity grows initially but then decays to 0.5 as expected for the statistical mixture (asymptotic behavior data not shown here).  When all three processes are present, the fidelity stabilizes at a value greater than 0.9.
(For fidelity of a density matrix $\rho$ with respect to the target state $|C_\alpha^+\rangle$, we use the definition: $F = \langle C_\alpha^+|\rho|C_\alpha^+\rangle$). Here we choose two-photon dissipation rate and the parity selection rate to be  
$\kappa_{\rm{2ph}} = 250 \kappa_{\rm{1ph}}, \kappa_{\rm{ps}} = 760 \kappa_{\rm{1ph}}$, consistent with the required inequality (1) above.

The master equation we have studied is an idealized ``cavity-only" system, whereas additional components will be required to realize the required baths and drives.  In the following section we propose a possible experimental implementation of the aforementioned stabilization scheme. Subsequently we will analyze the reduction of this system to the an effective model described by the single cavity master equation.

\section{Proposed Experimental Implementation}
\label{Experimental_Implementation}
\begin{figure}
\includegraphics[width = 0.5\textwidth]{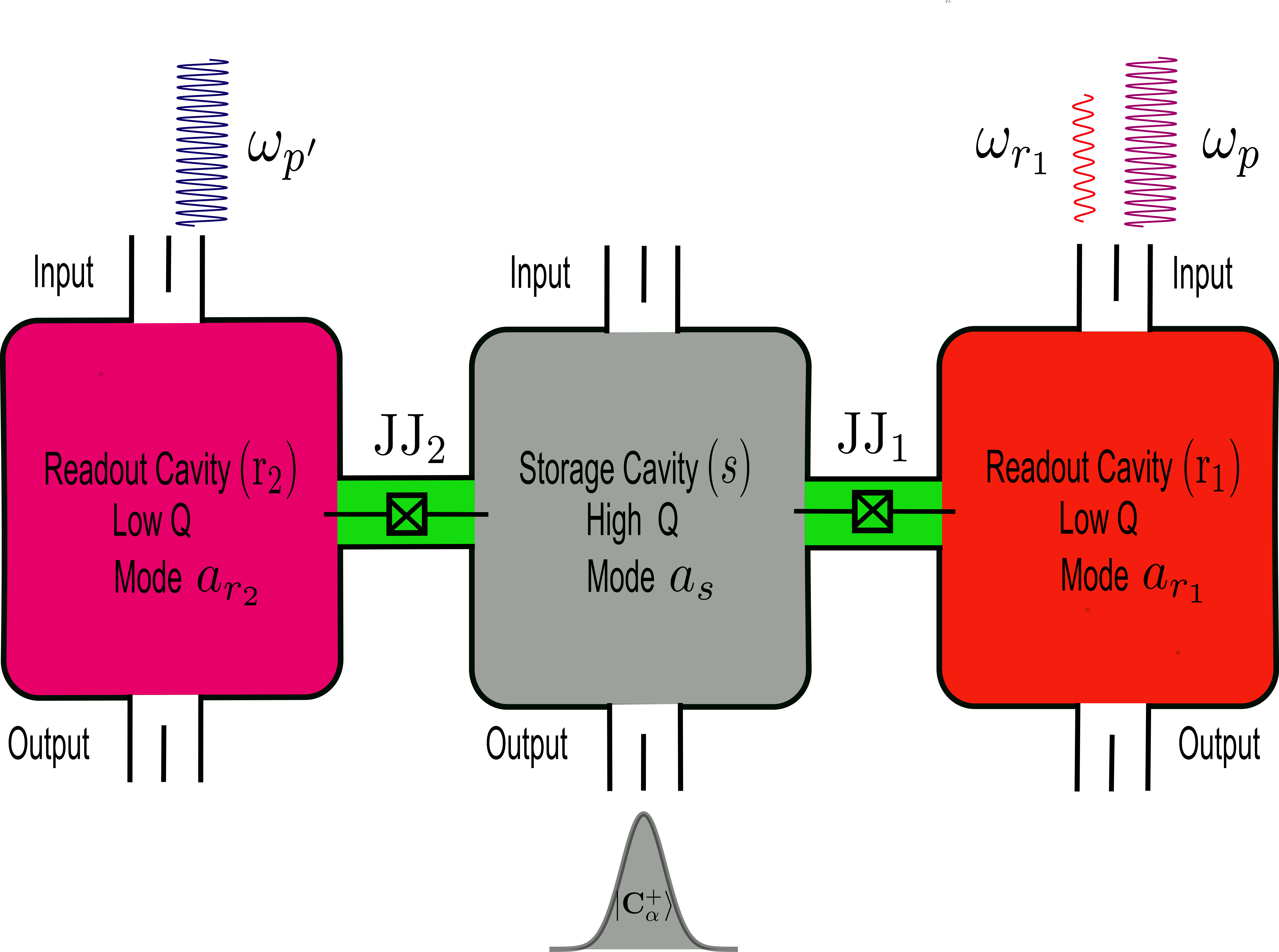}
\caption{\label{arch1} Schematic of experimental set-up realizing the stabilization scheme. Josephson junction $\rm{JJ}_1$ bridges the storage and readout cavity $r_1$. This, together with the stiff off-resonant pump at $\omega_p = 2\omega_s-\omega_{r_1}$, and the weak resonant drive at $\omega_{r_1}$ incident on $r_1$, gives rise to the two-photon drive and dissipation. Josephson junction $\rm{JJ}_2$ bridges the storage and readout cavity $r_2$ providing a nonlinear coupling between the modes $\mathbf{a_s}$ and low-Q mode $\mathbf{a_{r_2}}$. An off-resonant pump incident on $r_2$ at frequency $\omega_{p'} = (\omega_{r_2} - \omega_{s} - 2\tilde{n}\chi_{sr_2})/2$ gives rise to beam-splitter-like interaction between $\mathbf{a_s}$ and $\mathbf{a_{r_2}}$: $g_{\rm{ps}} e^{2i\omega_{p'}t}\mathbf{a_s}^\dagger \mathbf{a_{r_2}} + \rm{c.c.}$. This beam-splitter-like interaction acts conditioned on the mode $\mathbf{a_s}$ having $2\tilde{n}+1$ photons in the storage cavity. When the condition is realized, this interaction transfers one quantum of excitation from the $\mathbf{a_s}$-mode to the $\mathbf{a_{r_2}}$-mode, which is then lost irreversibly to the environment.}
\end{figure}

We propose a three-cavity two-junction architecture, where a high-Q cavity (referred to as storage cavity $s$) is linked by small transmission lines to two low-Q cavities, referred to as readout cavities $r_1$ and $r_2$ as shown in Fig.~\ref{arch1}. Each transmission line has an in-line embedded Josephson junction, which by virtue of the Josephson nonlinearity provides a nonlinear coupling between the storage and readout cavities. The single photon loss rate of the storage cavity is given by $\kappa_{\rm{1ph}}$, while that of the two readout cavities are given by $\kappa_{r_1}$ and $\kappa_{r_2}$ with the constraint: 
\begin{equation}
 \kappa_{\rm{1ph}}\ll\kappa_{r_1}, \kappa_{r_2}.
\end{equation}
The Hamiltonian of this device can be written as \cite{Nigg_Girvin_2012}: 
\begin{eqnarray}
\label{bbqham}
 \mathbf{H}_0 &=& \sum_k\hbar\omega_k\mathbf{a}_k^\dagger\mathbf{a}_k- E_{J_1}\big[\cos\Big(\frac{\mathbf{\Phi}_1}{\phi_0}\Big) + \frac{1}{2}\Big(\frac{\mathbf{\Phi}_1}{\phi_0}\Big)^2\big]\nonumber\\&& - E_{J_2}\big[\cos\Big(\frac{\mathbf{\Phi}_2}{\phi_0}\Big) + \frac{1}{2}\Big(\frac{\mathbf{\Phi}_2}{\phi_0}\Big)^2\big].
 \end{eqnarray}
Here $E_{J_{1,2}}$ are the Josephson energy for the two junctions,  $\omega_k$ are the bare frequencies of the modes $\mathbf{a}_k$,  $\phi_0 = \hbar/2e$ is the reduced flux quantum and $\mathbf{\Phi}_{1,2}$ is the flux through the Josephson junction linking readout cavity $r_{1,2}$ to storage cavity $s$. 

Here, only the fundamental modes of the three cavities are excited, annihilation operators (frequencies) of which are denoted respectively by $\mathbf{a_s} (\omega_{s}), \mathbf{a_{r_1}} (\omega_{r_1})$ and $\mathbf{a_{r_2}} (\omega_{r_2})$. The Josephson junctions ensure a nonlinear coupling of the  modes $\mathbf{a_s}$ and  $\mathbf{a_{r_1}}$ and similarly between the modes $\mathbf{a_s}$ and $\mathbf{a_{r_2}}$. This gives rise to self-Kerr and cross-Kerr interactions of the form: $-\frac{\chi_{ff}}{2}{\mathbf{f}^\dagger}^2\mathbf{f}^2$ and $-\chi_{fg}(\mathbf{f}^\dagger\mathbf{f})(\mathbf{g}^\dagger\mathbf{g})$, where $\mathbf{f}, \mathbf{g}$ correspond to the annihilation operators for the modes under consideration. Our stabilization scheme makes use of the following separation of time-scales (cf. Secs. \ref{twophotrealize}, \ref{parselrealize} for details): 
\begin{eqnarray}
\label{const1}
 \chi_{sr_1}\ll \kappa_{r_1} \text{\ and\ } \kappa_{r_2}\ll  \chi_{sr_2}.
\end{eqnarray}
This separation of time-scales can be engineered by appropriately choosing the participation ratios of the modes interacting through the junction nonlinearity. 

\begin{figure}
\includegraphics[width = 0.5\textwidth]{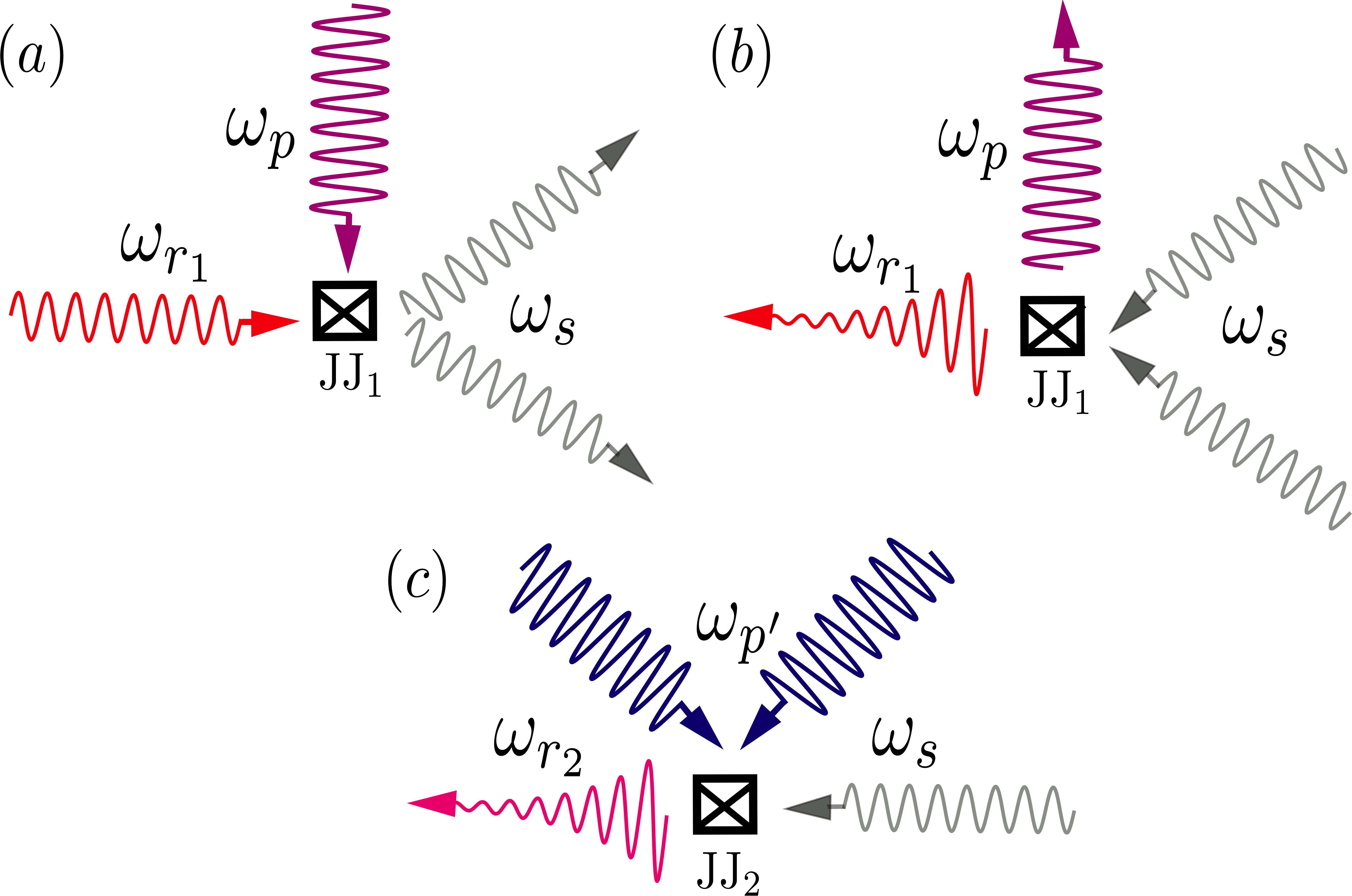}
\caption{\label{arch2} Scattering processes taking place through the nonlinear elements. (a) One photon in readout mode $\mathbf{a_{r_1}}$, together with one photon of pump at $\omega_p$ gets converted to two photons in mode $\mathbf{a_s}$, giving rise to the two-photon drive. (b) Two photons of the mode $\mathbf{a_s}$ are converted into one photon in pump mode at frequency $\omega_{p}$ and one photon in mode $\mathbf{a_{r_1}}$, which then irreversibly decays to the environment, giving rise to two-photon dissipation. (c) One photon in the mode 
$\mathbf{a_s}$, along with two photons in the pump with the adequate frequency $\omega_{p'}$, are converted conditionally into a photon in mode $\mathbf{a_{r_2}}$, which then irreversibly decays to the environment. This process occurs only when the number of photons in the storage cavity is $2\tilde{n}+1$, giving rise to the parity-selection mechanism. }
\end{figure}

\subsection{Realizing Two-Photon Process}
\label{twophotrealize}
We can engineer a non-linear interaction between the two modes $\mathbf{a_s}$ and $\mathbf{a_{r_1}}$ by means of a stiff (non-depleted), off-resonant pump incident on the readout cavity $r_1$. The frequency $\omega_p$ of the pump is chosen to be $\omega_p = 2\omega_{s} - \omega_{r_1}$. In addition, we drive the mode $\mathbf{a_{r_1}}$ with a weak {\it resonant} tone of amplitude $\epsilon_{r_1}$ and frequency $\omega_{r_1}$. Following the same kind of analysis as in~\cite{Nigg_Girvin_2012} and setting $\hbar = 1$ for the rest of this work, one can write the effective interaction Hamiltonian between the modes $\mathbf{a_s}$ and $\mathbf{a_{r_1}}$ as (see Fig.~\ref{arch2}): 
\begin{eqnarray}
\label{h2ph}
 \mathbf{H}_{\rm{sr_1}} &=& \omega_{s} \mathbf{a_s}^\dagger\mathbf{a_s} + \omega_{r_1} \mathbf{a_{r_1}}^\dagger\mathbf{a_{r_1}} + g_{\rm{2ph}}(\mathbf{a_s}^{\dagger 2}\mathbf{a_{r_1}} \nonumber\\&&+ \mathbf{a_s}^{2}\mathbf{a_{r_1}}^\dagger) - \epsilon_{r_1}(\mathbf{a_{r_1}} + \mathbf{a_{r_1}}^\dagger) \nonumber\\&&  - \frac{\chi_{ss}}{2}{\mathbf{a_s}^\dagger}^2\mathbf{a_s}^2 - \frac{\chi_{r_1r_1}}{2}{\mathbf{a_{r_1}}^\dagger}^2\mathbf{a_{r_1}}^2\nonumber\\&& -\chi_{sr_1}(\mathbf{a_s}^\dagger\mathbf{a_s})(\mathbf{a_{r_1}}^\dagger\mathbf{a_{r_1}}),
\end{eqnarray}
where we have assumed the nonlinear coupling $g_{\rm{2ph}}$ and drive amplitude $\epsilon_{r_1}$ to be real (phase of $g_{\rm{2ph}}$ is fixed by the phase of the stiff pump at $\omega_p$) and neglected nonlinearity higher than fourth order in mode amplitudes. In writing Eq. \eqref{h2ph}, we have also included self-Kerr and cross-Kerr interaction terms of the modes $\mathbf{a_s}, \mathbf{a_{r_1}}$ arising out of $\mathbf{H}_{0}$. As shown in~\cite{Mirrahimi_Devoret_2014}, 
\begin{equation}
\label{g2ph}
g_{\rm{2ph}} = \frac{\epsilon_p}{\omega_p-\omega_{r_1}}\chi_{sr_1}/2,
\end{equation}
where $\epsilon_p$ is the amplitude of the pump drive. For the rate inequalities given by (Eq. \eqref{const1}),  the Hamiltonian (Eq. \eqref{h2ph}), together with the decay of the low-Q mode $\mathbf{a_{r_1}}$, gives rise to the two-photon drive and dissipation of Eq. \eqref{w/oerror} (cf. \cite{Zaki} and Chap. 12 of \cite{Carmichael_2007} for details of calculation). 

\subsection{Realizing Parity Selection}
\label{parselrealize}
Next, we describe the interaction between the modes $\mathbf{a_s}$ and $\mathbf{a_{r_2}}$. We propose to engineer a beam-splitter-like interaction of the form $\mathbf{a_s}\mathbf{a_{r_2}}^\dagger + \mathbf{a_s}^\dagger \mathbf{a_{r_2}}$ {\it conditioned on the number of photons in the $\mathbf{a_s}$-mode being $2\tilde{n} + 1$}. This interaction has the effect that when the mode $\mathbf{a_s}$ has $2\tilde{n} + 1$ photons, a photon of the $\mathbf{a_s}$ mode is destroyed, in turn creating a photon in the mode $\mathbf{a_{r_2}}$, which is rapidly and irreversibly lost to the environment due to its low-Q nature of resonator $r_2$. This  state-selective beam-splitter interaction is generated by a stiff pump incident on the readout cavity $r_2$  at frequency $\omega_{p'} = (\omega_{r_2} - \omega_{s} - 2\tilde{n}\chi_{sr_2})/2$ (see below for more details). To realize the number-selectivity of this interaction, we need to work in the strong dispersive regime of the storage cavity. This ensures that the beam-splitter interaction becomes off-resonant when the number of photons in mode $\mathbf{a_s}$ is anything but $2\tilde{n}+1$. The  Hamiltonian describing the interaction between modes $\mathbf{a_s}$ and $\mathbf{a_{r_2}}$ is given by (see Fig. \ref{arch2}): 
\begin{eqnarray}
 \label{hcorr}
 \mathbf{H}_{\rm{sr_2}} &=& \omega_{s} \mathbf{a_s}^\dagger\mathbf{a_s} + \omega_{r_1} \mathbf{a_{r_2}}^\dagger\mathbf{a_{r_2}} + g_{\rm{ps}} \big(e^{2i\omega_{p'}t}\mathbf{a_s}^\dagger \mathbf{a_{r_2}}\nonumber\\&& + e^{-2i\omega_{p'}t}\mathbf{a_s}\mathbf{a_{r_2}}^\dagger\big) - \frac{\chi_{ss}}{2}{\mathbf{a_s}^\dagger}^2{\mathbf{a_s}}^2\\&& - \frac{\chi_{r_2r_2}}{2}{\mathbf{a_{r_2}}^\dagger}^2{\mathbf{a_{r_2}}}^2 -\chi_{sr_2}(\mathbf{a_s}^\dagger\mathbf{a_s})(\mathbf{a_{r_2}}^\dagger\mathbf{a_{r_2}})\nonumber,
\end{eqnarray}
where $g_{\rm{ps}}$ is the strength of the beam-splitter interaction fixed by the pump amplitude ($\epsilon_{p'}$) and is given by: 
\begin{equation}
\label{gpar}
g_{\rm{ps}} = \sqrt{\chi_{r_2r_2}\chi_{sr_2}}\Big|\frac{\epsilon_{p'}}{\omega_{p'} - \omega_{r_2}}\Big|^2.
\end{equation}
Due to the rate inequalities of Eq. \eqref{const1}, it suffices to keep only the cross-Kerr interaction $-\chi_{sr_2}(\mathbf{a_s}^\dagger\mathbf{a_s})(\mathbf{a_{r_2}}^\dagger\mathbf{a_{r_2}})$ for the calculation. The selectivity of the transition between the levels $|2\tilde{n} + 1\rangle_{a_s}\otimes|0\rangle_{a_{r_2}}$ and $|2\tilde{n}\rangle_{a_s}\otimes|1\rangle_{a_{r_2}}$ is ensured by detuning the frequency of the stiff pump ($\omega_{p'}$) from 
$(\omega_{r_2} - \omega_{s})/2$ by  $- \tilde{n}\chi_{sr_2}$. This leads to strong number selectivity
when $\chi_{sr_2}\gg g_{\rm{ps}}$. In addition, the cross-Kerr interaction has also to be stronger than the damping of the low-Q mode $\mathbf{a_{r_2}}$, i.e. $\chi_{sr_2}\gg\kappa_{r_2}$ so that the state-selectivity is not washed away by dissipation-induced level-broadening.  

Moving to the rotating frame $\mathbf{a_s}\rightarrow\mathbf{a_s}e^{-i\omega_{s}t}, \mathbf{a_{r_1}}\rightarrow\mathbf{a_{r_1}}e^{-i\omega_{r_1}t}, \mathbf{a_{r_2}}\rightarrow\mathbf{a_{r_2}}e^{-i\omega_{r_2}t + 2i\tilde{n}\chi_{sr_2}t}$, we can now write down the master equation for the density matrix ($\rho_{sr_1r_2}$) for the full 
three-mode model associated with $\mathbf{a_s},\mathbf{a_{r_1}}$ and $\mathbf{a_{r_2}}$:
\begin{eqnarray}
\label{3modemastereqn}
 \frac{d\rho_{sr_1r_2}}{dt} &=& -i\big[\mathbf{\bar{H}}_{\rm{2ph}} + \mathbf{H}_{\rm{ps}} + \mathbf{H}_{\rm{cross-Kerr}}, \rho_{sr_1r_2}\big]\nonumber \\&& + \big[\kappa_{r_1}{\cal D}(\mathbf{a_{r_1}}) + \kappa_{r_2}{\cal D}(\mathbf{a_{r_2}})\nonumber\\&& + \kappa_{\rm{1ph}}{\cal D}(\mathbf{a_s})\big]\rho_{sr_1r_2},
\end{eqnarray}
where 
\begin{eqnarray}
\label{hamcomp1}
\mathbf{\bar{H}}_{\rm{2ph}} &=& g_{\rm{2ph}}(\mathbf{a_s}^{\dagger 2}\mathbf{a_{r_1}} + \mathbf{a_s}^{2}\mathbf{a_{r_1}}^\dagger) - \epsilon_{r_1}(\mathbf{a_{r_1}} + \mathbf{a_{r_1}}^\dagger),\nonumber\\
 \mathbf{H}_{\rm{ps}} &=& g_{\rm{ps}}(\mathbf{a_s}\mathbf{a_{r_2}}^\dagger + \mathbf{a_s}^\dagger\mathbf{a_{r_2}}),\nonumber\\\mathbf{H}_{\rm{cross-Kerr}} &=& \chi_{sr_2}\big(2\tilde{n} - \mathbf{a_s}^\dagger \mathbf{a_s}\big)\mathbf{a_{r_2}}^\dagger\mathbf{a_{r_2}}.
\end{eqnarray}

\begin{figure}[!h]
\centering
\includegraphics[width = 0.5\textwidth]{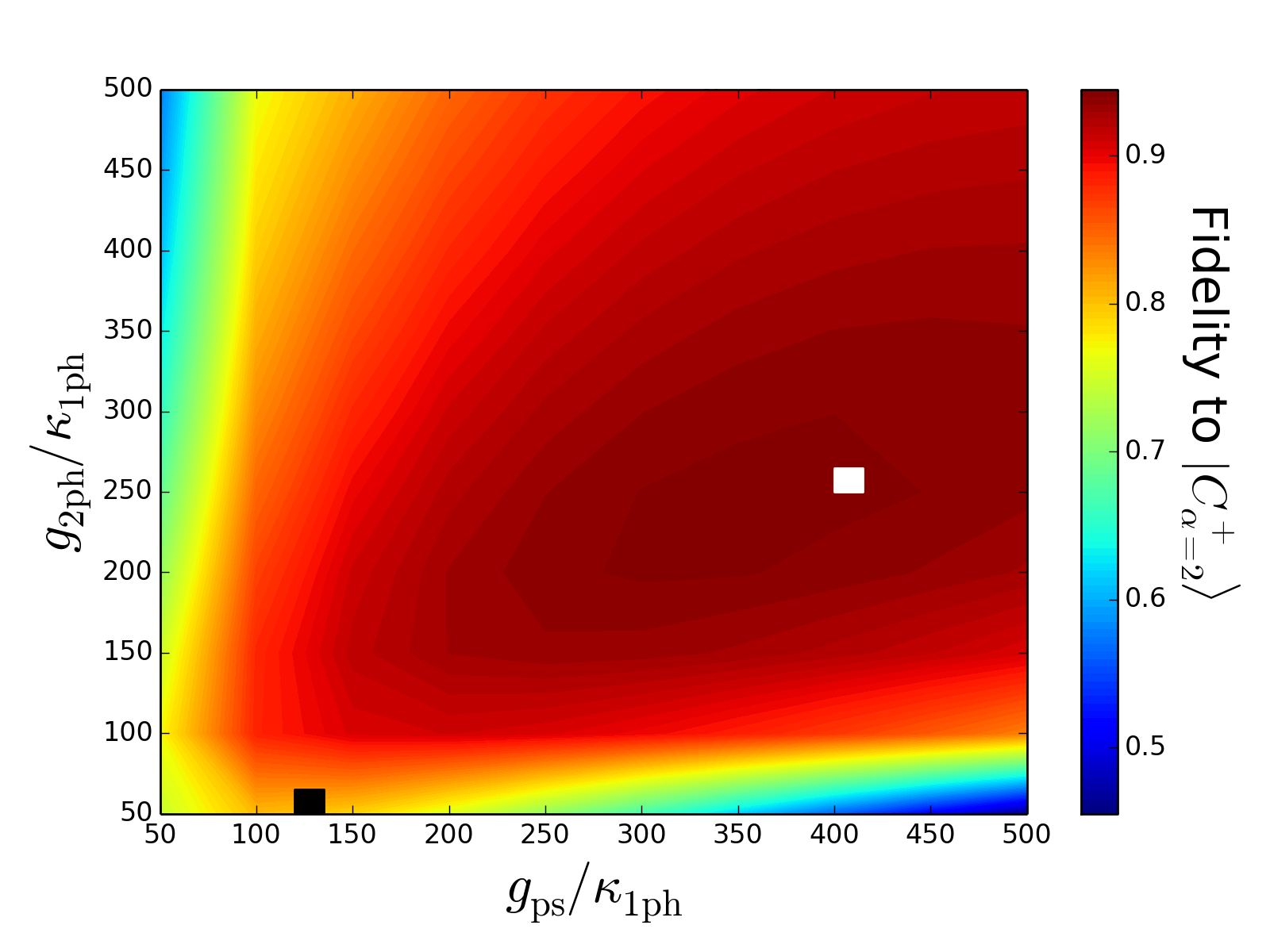}
\caption{\label{fid_varn_plot} Fidelity with respect to the target cat state, obtained by solving Eq. \eqref{3modemastereqn},versus parameters $g_{\rm{2ph}}/\kappa_{\rm{1ph}}, g_{\rm{ps}}/\kappa_{\rm{1ph}}$. We choose $\kappa_{r_1} = \kappa_{r_2} = 10^3\kappa_{\rm{1ph}}, \chi_{sr_2}=2.5\times10^4\kappa_{\rm{1ph}}$. The ratio of  $\epsilon_{r_1}$ and  $g_{\rm{2ph}}$ is chosen to be $4$ so that the target cat state is $|C_{\alpha=2}^+\rangle$. White square denotes the point of optimal fidelity, $\simeq  0.94$ for this choice of parameters ($g_{\rm{2ph}} = 250\kappa_{\rm{1ph}}, g_{\rm{ps}} = 400\kappa_{\rm{1ph}}$, cf. Fig. \ref{corrcomp2})). The black square is the point in the shown range of parameters where the adiabatic elimination of Sec. \ref{adel} works best.}
\end{figure}

We now present the numerical results obtained from solving numerically the above three-mode master equation. 
In Fig. (\ref{fid_varn_plot}) we plot the fidelity with respect to the target cat state ($|C_{\alpha = 2}^+\rangle$) upon variation of the parameters $g_{\rm{2ph}}/\kappa_{\rm{1ph}}$ and $g_{\rm{ps}}/\kappa_{\rm{1ph}}$.The choice of parameters is as follows: $\kappa_{r_1} = \kappa_{r_2} = 10^3\kappa_{\rm{1ph}}, \chi_{r_1s}=2.5\times10^4\kappa_{\rm{1ph}}$. The ratio $\epsilon_{r_1}/g_{\rm{2ph}}=4$, so that the target cat state is 
$|C_{\alpha=2}^+\rangle$. We see that for this choice of parameters, the optimal fidelity ($\sim 0.94$) is obtained for $g_{\rm{2ph}} = 250\kappa_{\rm{1ph}}, g_{\rm{ps}} = 400\kappa_{\rm{1ph}}$. The robustness of the scheme is indicated by the fact that for a large range of parameters, we find fidelities in excess of $90\%$. Note that $g_{\rm{2ph}}$ cannot be increased arbitrarily; due to the inequalities \eqref{const1}, \eqref{g2ph}, $g_{\rm{2ph}}\leq\kappa_{r_1}$. $g_{\rm{ps}}$ also is bounded, by $\sqrt{\chi_{r_2r_2}\chi_{sr_2}}$ (cf. Eq. \eqref{gpar}), which is much larger than $\kappa_{r_2}$. For both these variables, these bounds are not reached in our simulations. 

In the following subsection, we will show how the above three-mode master equation (Eq. \eqref{3modemastereqn}) can be reduced to the single-mode effective master equation (Eq. \eqref{werrorcorr}), with the two-photon dissipation and parity-selection rates given by Eqs. \eqref{kappa2ph}, \eqref{kcorr}.

\subsection{Elimination of Fast Dynamics}
\label{adel}
Due to the low-Q nature of the modes $\mathbf{a_{r_1}}$ and $\mathbf{a_{r_2}}$, we can eliminate their dynamics adiabatically to arrive at a reduced equation of motion for mode $\mathbf{a_s}$. Elimination of the $\mathbf{a_{r_1}}$ mode can be done following Chap. 12 of \cite{Carmichael_2007}. This gives rise to a two-photon dissipation rate: 
\begin{equation}
\label{kappa2ph}
\kappa_{\rm{2ph}} = \frac{4g_{\rm{2ph}}^2}{\kappa_{r_1}}.
\end{equation}

After eliminating the mode  $\mathbf{a_{r_1}}$, we proceed to eliminate the fast dynamics associated with the mode $\mathbf{a_{r_2}}$. In the rotating frame of the Hamiltonian $\mathbf{H}_{\rm{cross-Kerr}}$, the reduced master equation for the density matrix ($\rho_{sr_2}$) for the modes $\mathbf{a_s}, \mathbf{a_{r_2}}$ is given by:
\begin{eqnarray}
\label{2modemastereqn}
 \frac{d\rho_{sr_2}}{dt} &=& -i\big[i\big(\epsilon_{\rm{2ph}}\mathbf{a_s}^{\dagger2} - \epsilon_{\rm{2ph}}^*\mathbf{a_s}^2\big)\mathbf{\Pi}_{|0\rangle_{a_{r_2}}}, \rho_{sr_2}\big]\nonumber\\&& + \kappa_{\rm{2ph}}{\cal D}(\mathbf{a_s}^2\mathbf{\Pi}_{|0\rangle_{a_{r_2}}})\rho_{sr_2} + \mathbf{\cal L}_{sr_2}\rho_{sr_2},
\end{eqnarray}
where
\begin{eqnarray}
\label{rhoacmast}
  \mathbf{\cal L}_{sr_2}\rho_{sr_2} &=& -ig_{\rm{ps}}\sum_{j=0}^\infty\Big\{\big[\mathbf{\Pi}_{|2\tilde{n}+1-j\rangle_{a_s}\otimes|j\rangle_{a_{r_2}}}\mathbf{a_s}^\dagger \mathbf{a_{r_2}}\nonumber\\&& + \mathbf{a_s} \mathbf{a_{r_2}}^\dagger \mathbf{\Pi}_{|2\tilde{n}+1-j\rangle_{a_s}\otimes|j\rangle_{a_{r_2}}}, \rho_{sr_2}\big]\nonumber\\&& + \kappa_{r_2}{\cal D}(\mathbf{a_{r_2}}\mathbf{\Pi}_{|j\rangle_{a_s}})\rho_{sr_2}\nonumber\\&& + \kappa_{\rm{1ph}}{\cal D}(\mathbf{a_s}\mathbf{\Pi}_{|j\rangle_{a_{r_2}}})\rho_{sr_2}\Big\}
\end{eqnarray}
and $\mathbf{\Pi}_{|0\rangle_{a_{r_2}}} = |0\rangle_{a_{r_2}}{}_{a_{r_2}}\langle0|, \mathbf{\Pi}_{|2\tilde{n}+1-j\rangle_{a_s}\otimes|j\rangle_{a_{r_2}}} = |2\tilde{n}+1-j\rangle_{a_s}\otimes|j\rangle_{a_{r_2}}{}_{a_{r_2}}\langle j|\otimes{}_{a_s}\langle2\tilde{n}+1-j|$. In writing Eqs. \eqref{2modemastereqn}, \eqref{rhoacmast}, we have made use of the rotating wave approximation, assuming that $\chi_{sr_2}\gg\kappa_{r_2}, g_{\rm{ps}}$.
In principle, for $\kappa_{r_2}> g_{\rm{ps}}$, we can adiabatically eliminate the dynamics of the low-Q mode $\mathbf{a_{r_2}}$. However, a direct calculation from Eqs. \eqref{2modemastereqn}, \eqref{rhoacmast} is difficult since any level of the mode-$\mathbf{a_{r_2}}$ can be excited. Instead, we approximately calculate an effective rate of transition of the system from the state $|2\tilde{n}+1\rangle_{a_s}\otimes|0\rangle_{a_{r_2}}$ to the state $|2\tilde{n}\rangle_{a_s}\otimes|0\rangle_{a_{r_2}}$ via the state $|2\tilde{n}\rangle_{a_s}\otimes|1\rangle_{a_{r_2}}$. Note that since the $\mathbf{a_{r_2}}$-mode is low-Q and there is no drive resonant at $\omega_{r_2}$, $\mathbf{a_{r_2}}$ gets populated solely due to the interaction term of the form $\mathbf{a_s}\mathbf{a_{r_2}}^\dagger$ in Eq. \eqref{rhoacmast}.  Hence we can expand the two-mode density matrix $\rho_{sr_2}$ as: 
\begin{eqnarray}
 \rho_{sr_2} &=&\rho_{00}|0\rangle_{a_{r_2}}{}_{a_{r_2}}\langle 0| + \delta\big(\rho_{01}|0\rangle_{a_{r_2}}{}_{a_{r_2}}\langle 1|\nonumber\\&&+ \rho_{10}|1\rangle_{a_{r_2}}{}_{a_{r_2}}\langle 0|\big) + \delta^2\big(\rho_{11}|1\rangle_{a_{r_2}}{}_{a_{r_2}}\langle 1|\nonumber\\&& + \rho_{20}|2\rangle_{a_{r_2}}{}_{a_{r_2}}\langle 0|+ \rho_{02}|0\rangle_{a_{r_2}}{}_{a_{r_2}}\langle 2|\big)\nonumber\\&& + {\cal O}(\delta^3),
\end{eqnarray}
where $\rho_{ij}, i,j=0, 1, 2$ act on the Hilbert space of the $\mathbf{a_s}$-mode. The natural small parameter of expansion is $\delta = g_{\rm{ps}}/\kappa_{r_2}$ (for similar analysis, cf. \cite{Wiseman_Milburn_1993}). We will show that the short-lived states $\rho_{01}, \rho_{10}$ and $\rho_{11}$ can be adiabatically eliminated in favor of an effective dynamics of $\rho_{00}$. We will also see that $\rho_{20}, \rho_{02}$ can be dropped for a reduced dynamics in the sector of Hilbert space of $\mathbf{a_s}$ which is of interest to us: span of $\big\{|2\tilde{n}\rangle_{a_s}, |2\tilde{n}+1\rangle_{a_s}\big\}$.  For this calculation, we omit the two-photon drive/dissipation which acts only on $\rho_{00}$ and the single photon loss, the rate of which is much slower than the fast time-scale of the adiabatic elimination. These terms gives rise to a correction only in orders of ${\cal O}(\kappa_{\rm{1ph}}/\kappa_{r_2})$ and can be neglected. We will reinsert them at the end to get the final evolution of the reduced density matrix of mode $\mathbf{a_s}$. Thus, from Eqs. \eqref{2modemastereqn},\eqref{rhoacmast}, we can write down an equation of motion for $\rho_{ij}, i,j=0,1,2$ in dimensionless variable $\tau = \kappa_{r_2} t$: 
\begin{eqnarray}
 \label{adelcal}
 \frac{d\rho_{00}}{d\tau} &=& -i\delta^2\big(\mathbf{\Pi}_{|2\tilde{n}+1\rangle_{a_s}}\mathbf{a_s}^\dagger\rho_{10}-\rho_{01}\mathbf{a_s}\mathbf{\Pi}_{|2\tilde{n}+1\rangle_{a_s}}\big)\nonumber\\&& + \delta^2\sum_{n=0}^\infty\mathbf{\Pi}_{|n\rangle_{a_s}}\rho_{11}\mathbf{\Pi}_{|n\rangle_{a_s}},\nonumber\\\frac{d\rho_{11}}{d\tau} &=& -i\big(\mathbf{a_s}\mathbf{\Pi}_{|2\tilde{n}+1\rangle_{a_s}}\rho_{01}-\rho_{10}\mathbf{\Pi}_{|2\tilde{n}+1\rangle_{a_s}}\mathbf{a_s}^\dagger\big) - \rho_{11},\nonumber\\\frac{d\rho_{01}}{d\tau} &=& -i\big(\delta^2\mathbf{\Pi}_{|2\tilde{n}+1\rangle_{a_s}}\mathbf{a_s}^\dagger\rho_{11}-\rho_{00}\mathbf{\Pi}_{|2\tilde{n}+1\rangle_{a_s}}\mathbf{a_s}^\dagger\nonumber\\&& - \sqrt{2}\delta^2\rho_{02}\mathbf{a_s}\mathbf{\Pi}_{|2\tilde{n}+2\rangle_{a_s}}\big) - \frac{1}{2}\rho_{01},\nonumber\\\frac{d\rho_{10}}{d\tau} &=& -i\big(\mathbf{a_s}\mathbf{\Pi}_{|2\tilde{n}+1\rangle_{a_s}}\rho_{00} + \sqrt{2}\delta^2\mathbf{\Pi}_{|2\tilde{n}+2\rangle_{a_s}}\mathbf{a_s}^\dagger\rho_{20}\nonumber\\&&-\delta^2\rho_{11}\mathbf{a_s}\mathbf{\Pi}_{|2\tilde{n}+1\rangle_{a_s}}\big) - \frac{1}{2}\rho_{10},\nonumber\\\frac{d\rho_{20}}{d\tau} &=& -i\sqrt{2}\mathbf{a_s}\mathbf{\Pi}_{|2\tilde{n}+2\rangle_{a_s}}\rho_{10} - \rho_{20},\nonumber\\\frac{d\rho_{02}}{d\tau} &=& i\sqrt{2}\rho_{01}\mathbf{\Pi}_{|2\tilde{n}+2\rangle_{a_s}}\mathbf{a_s}^\dagger - \rho_{02}.
\end{eqnarray}

Define: 
\begin{equation}
 \rho_{ij}^{m} = {}_{a_s}\langle m|\rho_{ij}|m\rangle_{a_s},\  i,j = 0, 1, m = 2\tilde{n}, 2\tilde{n}+1,\nonumber
 \end{equation}
\begin{equation}
 \bar{\rho}_{ij} = {}_{a_s}\langle 2\tilde{n}|\rho_{ij}|2\tilde{n}+1\rangle_{a_s}, \ \bar{\bar{\rho}}_{ij} = {}_{a_s}\langle 2\tilde{n}+1|\rho_{ij}|2\tilde{n}\rangle_{a_s}.
\end{equation}
Then, from Eqn. \eqref{adelcal}, we can write down: 
\begin{eqnarray}
 \label{rho2n+1}
 \frac{d\rho_{00}^{2\tilde{n}+1}}{d\tau} &=& -i\delta^2\sqrt{2\tilde{n}+1}\big(\bar{\rho}_{10}-\bar{\bar{\rho}}_{01}\big) + \delta^2\rho_{11}^{2\tilde{n}+1}\nonumber\\\frac{d\rho_{11}^{2\tilde{n}+1}}{d\tau} &=& -\rho_{11}^{2\tilde{n}+1}\nonumber\\\frac{d\rho_{01}^{2\tilde{n}+1}}{d\tau} &=& -i\delta^2\sqrt{2\tilde{n}+1}\bar{\rho}_{11} - \frac{1}{2}\rho_{01}^{2\tilde{n}+1}\nonumber\\\frac{d\rho_{10}^{2\tilde{n}+1}}{d\tau} &=& i
 \delta^2\sqrt{2\tilde{n}+1}\bar{\bar{\rho}}_{11} - \frac{1}{2}\rho_{10}^{2\tilde{n}+1}.
\end{eqnarray}

We see that the dynamics of $\rho_{11}^{2\tilde{n}+1}, \rho_{01}^{2\tilde{n}+1}$ and $\rho_{10}^{2\tilde{n}+1}$ occur on a much faster time-scale  than $\rho_{00}^{2\tilde{n}+1}$ and thus, while performing adiabatic elimination, we can replace them by their steady-state values: 
\begin{eqnarray}
\label{ss1}
 \big[\rho_{11}^{2\tilde{n}+1}\big]_{s.s.} &=& 0,\  \big[\rho_{01}^{2\tilde{n}+1}\big]_{s.s.} = -2i\delta^2\sqrt{2\tilde{n}+1}\big[\bar{\rho}_{11}\big]_{s.s.},\nonumber\\&& \big[\rho_{10}^{2\tilde{n}+1}\big]_{s.s.} = 2i\delta^2\sqrt{2\tilde{n}+1}\big[\bar{\bar{\rho}}_{11}\big]_{s.s.}.
\end{eqnarray}
Similarly, we can write down the equation of motion for $\rho_{ij}^{2\tilde{n}}$:
\begin{eqnarray}
 \label{rho2n}
 \frac{d\rho_{00}^{2\tilde{n}}}{d\tau} &=& \delta^2\rho_{11}^{2\tilde{n}}\nonumber\\\frac{d\rho_{11}^{2\tilde{n}}}{d\tau} &=& i\sqrt{2\tilde{n}+1}\big(\bar{\rho}_{10}-\bar{\bar{\rho}}_{01}\big) - \rho_{11}^{2\tilde{n}}\nonumber\\\frac{d\rho_{01}^{2\tilde{n}}}{d\tau} &=& i\sqrt{2\tilde{n}+1}\bar{\rho}_{00} - \frac{1}{2}\rho_{01}^{2\tilde{n}}\nonumber\\\frac{d\rho_{10}^{2\tilde{n}}}{d\tau} &=& -i\sqrt{2\tilde{n}+1}\bar{\bar{\rho}}_{00} - \frac{1}{2}\rho_{10}^{2\tilde{n}}, 
\end{eqnarray}
steady-state solutions of which give us:
\begin{eqnarray}
\label{ss2}
 \big[\rho_{11}^{2\tilde{n}}\big]_{s.s.} &=& -i\sqrt{2\tilde{n}+1}\big(\big[\bar{\bar{\rho}}_{01}\big]_{s.s.}-\big[\bar{\rho}_{10}\big]_{s.s.}\big),\nonumber\\ 
 \big[\rho_{01}^{2\tilde{n}}\big]_{s.s.} &=& 2i\sqrt{2\tilde{n}+1}\big[\bar{\rho}_{00}\big]_{s.s.},\nonumber\\  \big[\rho_{10}^{2\tilde{n}}\big]_{s.s.} &=& -2i\sqrt{2\tilde{n}+1}\big[\bar{\bar{\rho}}_{00}\big]_{s.s.}.
\end{eqnarray}
Using Eqns. \eqref{rho2n+1}, \eqref{ss1}, \eqref{rho2n},  \eqref{ss2}, we can write down equations of motion for $\rho_{00}^{2\tilde{n}}$ and $\rho_{00}^{2\tilde{n}+1}$: 
\begin{eqnarray}
 \label{rho00}
 \frac{d\rho_{00}^{2\tilde{n}+1}}{d\tau} &=& -i\delta^2\sqrt{2\tilde{n}+1}\big(\big[\bar{\rho}_{10}\big]_{s.s.}-\big[\bar{\bar{\rho}}_{01}\big]_{s.s.}\big),\nonumber\\\frac{d\rho_{00}^{2\tilde{n}}}{d\tau} &=& i\delta^2\sqrt{2\tilde{n}+1}\big(\big[\bar{\rho}_{10}\big]_{s.s.}-\big[\bar{\bar{\rho}}_{01}\big]_{s.s.}\big).
\end{eqnarray}
Note that $\frac{d\rho_{00}^{2\tilde{n}+1}}{d\tau} + \frac{d\rho_{00}^{2\tilde{n}}}{d\tau} = 0$, which signifies that the population of the state $|2\tilde{n}+1\rangle_{a_s}\otimes|0\rangle_{a_{r_2}}$ does indeed decay to  $|2\tilde{n}\rangle_{a_s}\otimes|0\rangle_{a_{r_2}}$. To complete the analysis and get an explicit form of the rate of population transfer, we write down the equation of motion for $\bar{\rho}_{10}, \bar{\bar{\rho}}_{01}$: 
\begin{eqnarray}
 \frac{d\bar{\rho}_{10}}{d\tau} &=& -i\sqrt{2\tilde{n}+1}\big(\rho_{00}^{2\tilde{n}+1}-\delta^2\rho_{11}^{2\tilde{n}}\big) - \frac{1}{2}\bar{\rho}_{10}\nonumber\\
 \frac{d\bar{\bar{\rho}}_{01}}{d\tau} &=& i\sqrt{2\tilde{n}+1}\big(\rho_{00}^{2\tilde{n}+1}-\delta^2\rho_{11}^{2\tilde{n}}\big) - \frac{1}{2}\bar{\bar{\rho}}_{01},\nonumber
\end{eqnarray}
steady state solutions of which are: 
\begin{eqnarray}
\label{ss3}
 \big[\bar{\rho}_{10}\big]_{s.s.}&=& - \big[\bar{\bar{\rho}}_{01}\big]_{s.s.}\nonumber\\  &=& -2i\sqrt{2\tilde{n}+1}\big(\rho_{00}^{2\tilde{n}+1}-\delta^2\rho_{11}^{2\tilde{n}}\big).
\end{eqnarray}
\begin{figure}[!h]
\centering
\includegraphics[width = 0.5\textwidth]{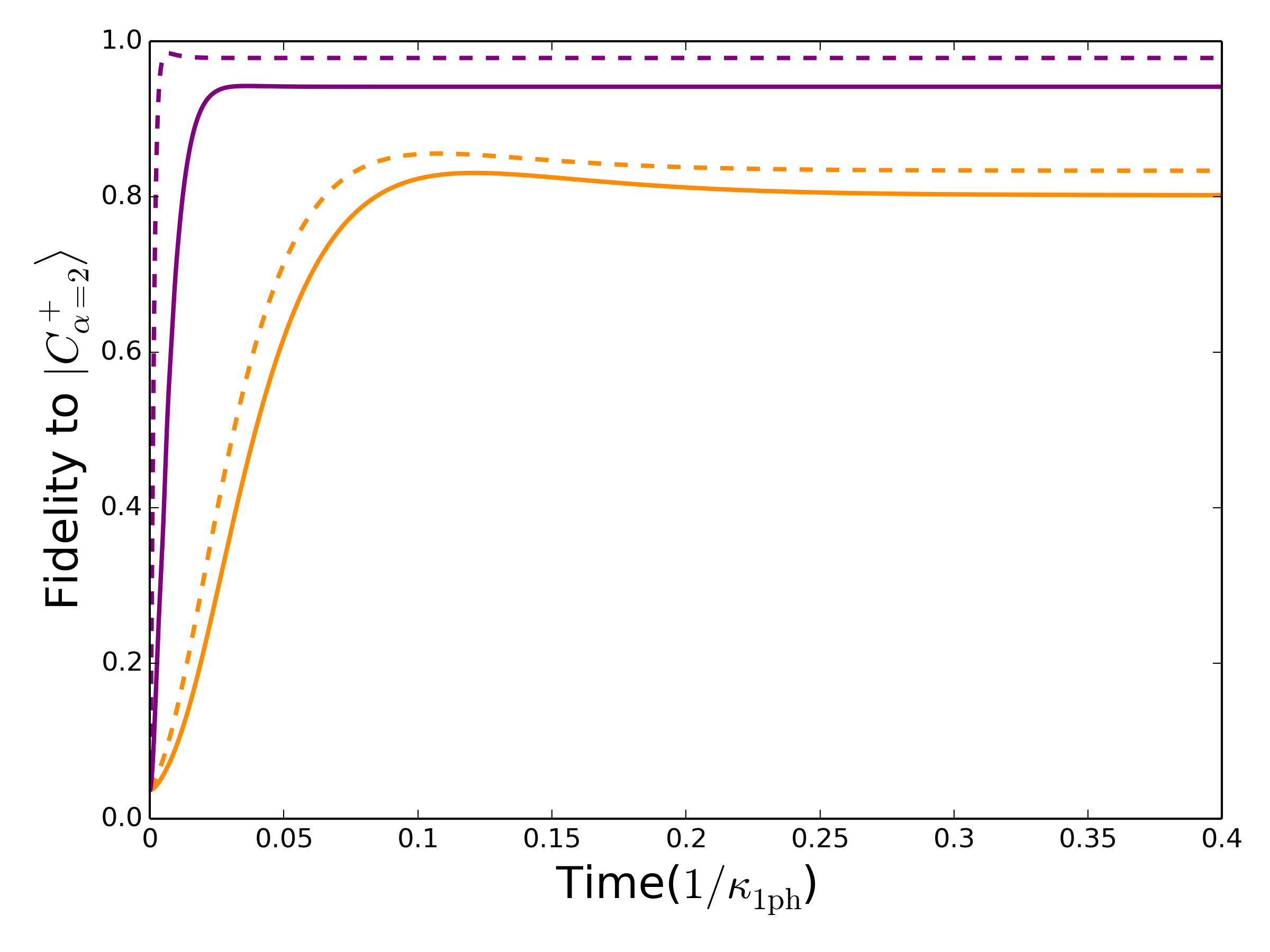}
\caption{\label{adelcomp} Comparison between the evolution of fidelities for the full three-mode master equation (Eq. \eqref{3modemastereqn}) in solid lines, and that obtained from the reduced dynamics (Eqs. \eqref{werrorcorr}, \eqref{kcorr}) in dashed lines. The two sets of parameters are chosen from Fig. \ref{fid_varn_plot}: the white square (purple curves) corresponding to  $g_{\rm{ps}} = 400\kappa_{\rm{1ph}}, g_{\rm{2ph}} = 250\kappa_{\rm{1ph}}$ and $\epsilon_{r_1} = 1000\kappa_{\rm{1ph}}$ and the black square (orange curves) corresponding to  $g_{\rm{ps}} = 120\kappa_{\rm{1ph}}, g_{\rm{2ph}} = 50\kappa_{\rm{1ph}}$ and $\epsilon_{r_1} = 200\kappa_{\rm{1ph}}$. For both sets of curves, $\chi_{sr_2} = 2.5\times10^4\kappa_{\rm{1ph}}, \kappa_{r_2} = \kappa_{r_1} = 1000\kappa_{\rm{1ph}}$ and the target state is $|C_{\alpha = 2}^+\rangle$. The model-reduction  (Eq. \eqref{werrorcorr}) approaches the full three-mode master equation (Eq. \eqref{3modemastereqn}) as the adiabatic approximation ($g_{\rm{2ph}}/\kappa_{r_1}\ll1, g_{\rm{ps}}/\kappa_{r_2}\ll1$) and the rotating rotating wave approximation  ($g_{\rm{ps}}/\chi_{sr_2}\ll1, \kappa_{r_2}/\chi_{sr_2}\ll1$) become more and more accurate. }
\end{figure}

Using Eqs. \eqref{ss2}, \eqref{rho00}, \eqref{ss3} and some tedious algebra, we have (in dimensional variables): 
\begin{equation}
 \frac{d\rho_{00}^{2\tilde{n}+1}}{dt} = -\kappa_{\rm{ps}}\rho_{00}^{2\tilde{n}+1},\hspace{0.5cm}\frac{d\rho_{00}^{2\tilde{n}}}{dt} = \kappa_{\rm{ps}}\rho_{00}^{2\tilde{n}+1},
\end{equation}
where \begin{equation}
\label{kcorr}
       \kappa_{\rm{ps}} = \frac{4\delta^2(2\tilde{n}+1)}{1+4\delta^2(2\tilde{n}+1)}\kappa_{r_2}.
      \end{equation}
Thus we have indeed derived an effective dynamics for the reduced density matrix of the storage mode: $\rho = \rm{Tr}_{a_{r_2}}\big[\rho_{sr_2}\big]$ as given by Eq. \eqref{werrorcorr} of Sec. \ref{Outline_Stabilization_Scheme} with $\kappa_{\rm{ps}}$ given by Eq. \eqref{kcorr}. 

The key requirements for the above model reduction are the validity of the adiabatic approximation ($g_{\rm{2ph}}/\kappa_{r_1}\ll1, g_{\rm{ps}}/\kappa_{r_2}\ll1$) and the rotating wave approximation ($g_{\rm{ps}}/\chi_{sr_2}\ll1, \kappa_{r_2}/\chi_{sr_2}\ll1$). In Fig. \ref{adelcomp},	 we compare the validity of the model-reduction for two choice of parameters (cf. Fig. \ref{fid_varn_plot}): the white square (purple curves) corresponding to  $g_{\rm{ps}} = 400\kappa_{\rm{1ph}}, g_{\rm{2ph}} = 250\kappa_{\rm{1ph}}$ and $\epsilon_{r_1} = 1000\kappa_{\rm{1ph}}$ and the black square (orange curves) corresponding to  $g_{\rm{ps}} = 120\kappa_{\rm{1ph}}, g_{\rm{2ph}} = 50\kappa_{\rm{1ph}}$ and $\epsilon_{r_1} = 200\kappa_{\rm{1ph}}$. 
For both sets of curves, $\chi_{sr_2} = 2.5\times10^4\kappa_{\rm{1ph}}, \kappa_{r_2} = \kappa_{r_1} = 1000\kappa_{\rm{1ph}}$ and the target state is $|C_{\alpha = 2}^+\rangle$. 
The model-reduction  (Eq. \eqref{werrorcorr}) approaches the full three-mode master equation (Eq. \eqref{3modemastereqn}) as the adiabatic approximation ($g_{\rm{2ph}}/\kappa_{r_1}\ll1, g_{\rm{ps}}/\kappa_{r_2}\ll1$) and the rotating rotating wave approximation  ($g_{\rm{ps}}/\chi_{sr_2}\ll1, \kappa_{r_2}/\chi_{sr_2}\ll1$) become more and more accurate.

\section{Conclusions}
\label{concl}
Following recent advances in the production of non-classical states of light, we have proposed a scheme to prepare, and protect against decoherence, Schr\"{o}dinger cat states of given photon number parity. Relying only on the application of continuous-wave drives of fixed but carefully chosen frequencies, we are able to engineer an effective Hamiltonian and dissipation which stabilizes such states. The scheme is independent of the phase of the drives and  appears to be robust with respect to the choice of their amplitudes. Numerical simulations illustrate that the required parameters are within reach of the ongoing experiments in the field of quantum superconducting circuits. Such a stabilized source of Schr\"{o}dinger cat states is a valuable system component that could be integrated in existing quantum information processing schemes based only on linear optical scattering elements and amplifiers.

Discussions with Pierre Rouchon and Benjamin Huard are gratefully acknowledged. The work was supported by NSF grant no. ECCS 1068642 and U.S. Army Research Office W911NF-09-1-0514. A.R. and M.M. acknowledge the support of Idex ANR-10-IDEX-0001-02 PSL*.

\bibliography{references}
\end{document}